\documentclass[preprint,amssymb,nofootinbib,amsmath,11pt]{JHEP3}

\JHEPspecialurl{http://jhep.sissa.it/JOURNAL/JHEP3.tar.gz}

\usepackage{epsfig,multicol,amsmath}

\usepackage{epsfig,multicol}

\def\beq{\begin{equation}}
\def\eeq{\end{equation}}
\def\bea{\begin{eqnarray}}
\def\eea{\end{eqnarray}}

\def\eq#1{{Eq.~(\ref{#1})}}
\def\fig#1{{Fig.~\ref{#1}}}

\newcommand{\Lb}{\left(}
\newcommand{\Rb}{\right)}

\parskip=\itemsep               

\newcommand{\h}{\frac{1}{2}}

\newcommand{\A}{{\cal A}}

\title{\LARGE \bf Soft interaction at high energy and   N=4 SYM }
\author{\large  E. ~Levin$^{a}$,\, and \,\,I.~Potashnikova$^{b}$  \\
a)  \,Department of Particle Physics, School of Physics and Astronomy\\
Raymond and Beverly Sackler
 Faculty
of Exact Science\\  Tel Aviv University, Tel Aviv, 69978, Israel \\~~\\
E-mail address:\email{leving@post.tau.ac.il}\\
b)\,Departamento de F\'\i sica
y Centro de Estudios
Subat\'omicos,\\ Universidad T\'ecnica
Federico Santa Mar\'\i a, Avda. Espa\~na 1680,\\
Casilla 110-V, Valpara\'\i so, Chile \\
~~\\E-mail address:\email{irina.potashnikova@usm.cl}}

\abstract{ In this paper we show that the N=4 SYM total cross section violates the Froissart theorem, and in the huge range of energy this cross section is proportional to $s^{1/3}$. The graviton reggeization will change this increase to the normal logarithmic behavior $\sigma \propto \ln^2 s$.  However,  we demonstrated that this happens at  ultra high energy, much higher than the LHC energy. In the region of accessible energy we need to assume that there is a  different source for the total cross section, with the value of the cross section about 40 mb.  With this assumption we successfully describe $\sigma_{tot}, \sigma_{el}$ and $\sigma_{diff}$
for the accessible range of energy from the fixed target Fermilab to the Tevatron energies. It turns out that  the N=4 SYM mechanism can be  responsible only for a small part of the inelastic cross section for this energy region (about $2 mb$).
However, at the LHC energy the N=4 SYM theory can describe the multiparticle production with  $\sigma_{in} \approx 30\,mb$.
The second surprise is the fact that the total cross section and the diffraction cross section can increase considerably from the Tevatron to the LHC energy. The bad description of $B_{el}$ gives the strong argument that the non N=4 SYM background should depend on energy. We believe that we have a dilemma: to find a new mechanism for the inelastic production in the framework of N=4 SYM other than  the reggeized graviton interaction, or to accept that N=4 SYM is irrelevant to any  experimental data that has been  measured  before the LHC era.
}


\preprint{ TAUP 2891/08 \\
\today}
\begin{document}

\numberwithin{equation}{section}

\section{Introduction}
At the moment N=4  SYM is the unique theory which allows us to study theoretically the regime of the strong coupling constant \cite{AdS-CFT} .  Therefore, in principle, considering the high energy scattering
amplitude in N=4 SYM, we can guess which physics phenomena could be
important in QCD, in the limit of the strong coupling. The attractive feature of this theory, is that N=4 SYM with small coupling leads to normal QCD like physics (see Refs.
\cite{POST,BFKL4})  with OPE  and linear equations for DIS as well as the BFKL equation for the high energy amplitude.
The high energy amplitude reaches the unitarity limit: black disc regime, in which half of the cross section stems from the elastic scattering and half relates to the processes of the multiparticle production.

However, the physical picture  in the strong coupling region turns out
to be completely different \cite{MHI,BST1,BST2,BST2,COCO,BEPI,LMKS}, in the sense that there are no
processes of the multiparticle production
 in this region, and the main contribution stems from elastic and quasi-elastic (
diffractive)  processes when the target (proton) either remains
intact, or is slightly excited. Such a picture not  only contradicts
the QCD expectations \cite{GLR,MUQI,MV,BFKL,BK,JIMWLK}, but also contradicts available experimental data.

On the other hand, the main success of N=4 SYM  has been achieved in the description of the multiparticle system such as the quark-gluon plasma and/or the multiparticle system at fixed temperature \cite{KSS,HKKKY,MUN4,PLASMAN4}.
Therefore, we face a controversial situation: we know a lot about something that cannot be produced.

The goal of this paper is to  evaluate the scale of the disaster, comparing the predictions of the N=4  SYM with the experimental data. We claim that at least   half of the total cross section at the Tevatron  energy has to stem from a different source than the N=4  SYM.

Before discussing predictions of the N=4 SYM for high energy scattering, we  would like to draw  the reader's attention that there exists  two different regions  of energy that we have to consider in N=4 SYM: $ (2/\sqrt{\lambda})\,\alpha' s  <1$ and  $(2/\sqrt{\lambda})\,\alpha' s > 1$ ($ \lambda =\,4 \pi g_s N_c$ where $g_s$ is the string  coupling and $N_c$ is the number of colors). In the first region, the multiparticle production has a very small cross section, and it can be neglected. However, in the second region the graviton reggeization leads to the inelastic cross section that
is rather large, and at ultra high energies the scattering amplitude reveals all of the typical features of the black disc regime:
$\sigma_{el} = \sigma_{tot}/2$ and $ \sigma_{in} = \sigma_{tot}/2$.

Therefore, the formulation of the main result of this paper is the following: at the accessible energies the amplitude is in the first region, and at least   half of the total cross section at the Tevatron  energy has to stem from a different source than the N=4  SYM. However, at the LHC energy the N=4 SYM mechanism can be responsible for about 2/3 of the total cross section  and, perhaps, at the LHC  the final states will be produced with the typical properties of the N=4 SYM.

\section{High energy Scattering in N=4 SYM}
\subsection{Eikonal formula}
The main contribution to the scattering amplitude at high energy  in
N=4 SYM, stems from the exchange of the graviton\footnote{Actually,
the graviton in this theory is reggeized \cite{BST1}, but it is easy
to take this effect into account (see Refs. \cite{BST1,BST3,MHI})  and \eq{EQ} below.}
. The  formula for this exchange has been written in
Ref.\cite{BST2,COCO,LMKS}.
In $ AdS_5 =AdS_{d + 1}$ space this amplitude has the following form (see \fig{oge})

\beq \label{N41}
A_{1GE}(s,b; z,z')\,\,=\,\,g^2_s\,\,T_{\mu\nu}\Lb p_1,p_2\Rb G_{\mu \nu \mu^{\,\prime} \nu^{\,\prime}}\Lb u\Rb\,T_{\mu^{\,\prime}\nu^{\,\prime}}\Lb p_1,p_2\Rb\,\, \xrightarrow{s\gg \mu^2} g^2_s \, s^2 z^2z'^2\,G_3\Lb u \Rb
\eeq

\FIGURE[h]{\begin{minipage}{50mm}
\centerline{\epsfig{file=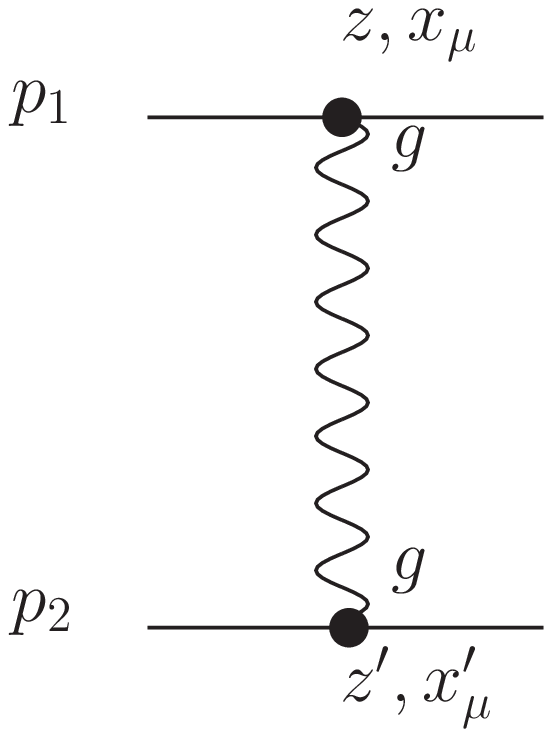,width=40mm}}
\caption{The one graviton (1GE) exchange.  }
\end{minipage}
\label{oge}}
where $T_{\mu,\nu}$ is the energy-momentum tensor, and $G$ is the
propagator of the massless graviton. The last expression in
\eq{N41}, reflects the fact that for high energies, $T_{\mu,\nu} =
p_{1,\mu} p_{1,\nu}$ and at high energies the momentum transferred $q^2 \,\to q^2_{\perp}$ which led to  $G_3\Lb u\Rb $ (see  Refs.\cite{BST2,LMKS}).  In $AdS_5$  the metric has  the following form
\beq \label{N44}
d s^2\,\,=\,\,\frac{L^2}{z^2}\,\Lb \,d z^2\,\,+\,\,\sum^d_{i=1}  d x^2_i \Rb\,=\,\frac{L^2}{z^2}\,\Lb \,d z^2\,+\,d \vec{x}^2 \Rb
\eeq
and $u$ is a new variable which is equal to

\beq \label{u}
u\,\,=\,\,\frac{ (z - z')^2 + (\vec{x} - \vec{x}')^2}{ 2 \,z\,z'}\,\,=\,\,\frac{ (z - z')^2 + b^2}{ 2 \,z\,z'}\,\,
\eeq

~

and
\beq \label{G}
G_{3}\Lb u \Rb \,\,=\,\,\frac{1}{4 \pi}\,\frac{1}{\left\{ 1 + u + \sqrt{u (u + 2)}\right\}^2\,\sqrt{u (u + 2)}}
\eeq
where $b$ is the impact parameter (see \fig{oge}).

As one can see from \eq{N41} the one graviton exchange amplitude is real. As has been discussed \cite{BST1} the graviton reggeization leads to  a small imaginary part, and the amplitude  can be re-written in the form \cite{BST1,LMKS}

\beq \label{N42}
\tilde{A}_{1GE}(s,b; z,z')\,\,\equiv\,\,\frac{A_{1GE}(s,b; z,z')}{s}\,\,=\,\, g^2_s \,\Lb 1 + i\rho\Rb\, s\, z\,z'\,G_3\Lb u \Rb
\eeq
where $\rho \,=\,2/\sqrt{\lambda}\,\ll\,1$. $\tilde{A}_{1GE}$ steeply increases with energy $s$ and has to be unitarized using the eikonal formula \cite{BST2,BST3,LMKS}
\beq \label{N43}
A_{eikonal}\Lb s,b;z,z'\Rb \,\,=\,\,i \Lb\,1\,\,\,-\,\,\,\exp
\Lb i\,   \tilde{A}_{1GE}\Lb s, b; \eq{N42}\Rb\Rb\Rb
 \eeq
In Ref. \cite{LMKS} it was argued that AdS/CFT correspondence leads to the corrections to \eq{N43} which are small
$(\propto 2/\sqrt{\lambda})$. The unitarity constraints for \eq{N43} has the form
\beq \label{N45}
2\,\mbox{Im} \A_{eikonal}\Lb s,b; z,z'\Rb\,\,\,=\,\,\,| A_{eikonal}\Lb s,b; z,z'\Rb|^2\,\,\,+\,\,{\cal O}\Lb \frac{2}{\sqrt{\lambda}}\Rb
\eeq

\FIGURE[h]{
\centerline{\epsfig{file=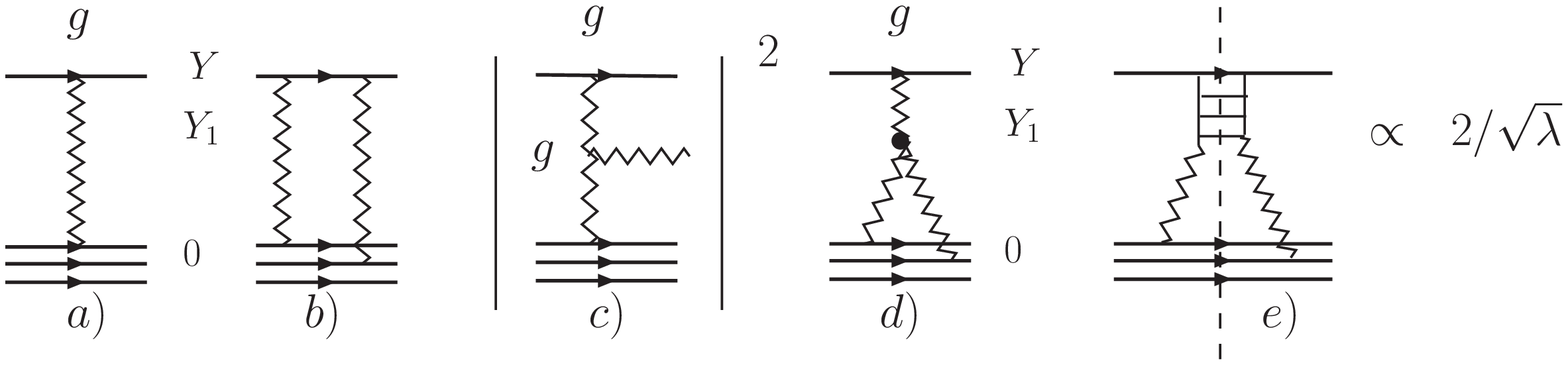,width=140mm,height=30mm}}
\caption{The diagrams for nucleon-nucleon interaction in N=4 SYM. \fig{din4}-a and \fig{din4}-b show the exchange of one and two gravitons that are included in the eikonal formula of \eq{N43}, while other diagrams give the examples of corrections to the eikonal formula.
\label{din4} }}
The eikonal formula of \eq{N43} as well as the unitarity constraint of \eq{N45} are illustrated in \fig{din4}.
One can see that the diagrams shown in this figure have the following contributions:
\beq \label{DIN4}
A\Lb \fig{din4}-a\Rb\,\propto\,g^2_s \,s\,\,\approx \,\frac{s}{N^2_c}\,;\,\,\,\,\,\,\,\,\,\,\,\,\,\,\,\,
A\Lb \fig{din4}-b\Rb\,\propto\,\Lb g^2_s \,s \Rb^2 \,\,\approx \,\Lb \frac{s}{N^2_c} \Rb^2\,;
\eeq
$$
A\Lb \fig{din4}-c\Rb\,\propto\, g^2_s\,\Lb g^2_s \,s \Rb^2 \,\,\approx \,\frac{1}{N^2_c}\Lb \frac{s}{N^2_c} \Rb^2\,;\,\,\,\,\,\,\,\,
A\Lb \fig{din4}-d \,\, \mbox{and} \,\, \fig{din4}-e\Rb\,\propto\,  \frac{2}{\sqrt{\lambda} }\,\Lb \frac{s}{N^2_c} \Rb^2.
$$
Therefore, the contributions that lead to a violation of the eikonal formula are small, at least as small as $2/\sqrt{\lambda}$. It is interesting to notice that actually they stem from the processes of the diffraction dissociation (see \fig{din4}-e rather than from the processes of the multiparticle productions (see \fig{din4}-c).

\eq{N42} provides the simple method to take into account the reggeization of the graviton, in order to understand the main property of the scattering amplitude. However in our description of the experimental data, we will use the exact form of the amplitude for the exchange of the reggeized graviton (see Refs. \cite{BST1, LMKS}), namely,
\beq \label{EQ}
\tilde{A}_{1GE}(s,b; z,z')\,\,=\,\,g^2_s \,( 1 + i \rho)\,\,\frac{1}{4 \pi}\,\frac{\,\Lb z\,z' s\Rb^{1 - \rho}}{\sqrt{u (u + 2)}}\,
\sqrt{\frac{\rho}{\pi\,\ln\Lb s \,z\,z'\Rb}}\,\exp \Lb - \frac{\ln^2\Lb 1 + u + \sqrt{u (u + 2 )}\Rb}{\rho\,\ln\Lb s\,z\,z'\Rb}\Rb
\eeq
\eq{EQ}  gives the description of one reggeized graviton in the limit $s \to \infty$ with $\lambda \,\gg\,1$ while the simple formula of \eq{N42} describes the one graviton exchange for $\lambda \to \infty$ but $s \gg 1/\alpha'$.

\subsection{Nucleon-nucleon high energy amplitude}
Discussing the hadron interaction at high energy, we need to specify the
correct degrees of freedom that diagonalize the interaction matrix. We assume that a nucleon consists of $N_c$ quarks  ($N_c$ colorless dipoles) that interact with each other with the eikonal formula of \eq{N43}, namely,
\bea\label{N46}
A_{NN}\Lb s,b\Rb \,\,\,&=&\,\,\,\int\,d z\,d z'\,\prod^{N_c}_{i=1}\, d^2 r_i \prod\,|\Psi\Lb r_i,z\Rb]^2\,\,
\prod^{N_c}_{i=1}\, d^2 r'_i \prod\,|\Psi\Lb r_i,z'\Rb]^2\,\nonumber\\
 &\times &\,\,\,i\,\Lb 1 \,\,-\,\,\,\,\,\exp
\Lb i\,   N^2_c\,\tilde{A}_{1GE}\Lb s, b; z,z' | \eq{N42}\Rb\Rb\Rb \nonumber \\
 &=& \,\,i\,\int\,d z\,d z'\,\Phi\Lb z\Rb\,\Phi\Lb z'\Rb\,\,\Lb 1\,\,-\,\,\exp\Lb i\,g^2 N^2_c\,( 1 + i \rho)\,z^2\,z'^2\,G_3\Lb u\Rb\Rb\Rb
\eea
where
\beq \label{PHI}
\Phi\Lb z\Rb\,\,=\,\,\int \, d^2 r \prod\,|\Psi\Lb r,z\Rb|^2\,\,
\eeq
and $\rho = 2/\sqrt{\lambda}$.

In \eq{N46} the only unknown ingredient is $\Psi\Lb r_i,z\Rb$.
 We can reconstruct  this wave function
 using the Witten formula \cite{WIT}, namely,

\bea \label{DP3}
&& \Psi \Lb r,  z\Rb\,\,\,=\\
&&\frac{\Gamma\Lb \Delta\Rb}{\pi\,\Gamma\Lb \Delta - 1\Rb}\,\,\int \,d^2 r' \,
\Lb \frac{z}{z^2\,\,+\,\,( \vec{r}\,-\,\vec{r}')^2}\Rb^{\Delta}\,\, \Psi\Lb r'\Rb
\,\,\,\mbox{with}\,\,\,\,\Delta_{\pm}\,\,=\,\,\h \Lb d\,\,\pm\,\,\sqrt{d^2 + 4\, m^2}\Rb \nonumber
\eea
where $\Psi\Lb r'\Rb$ is the wave function of the dipole inside the proton on the boundary. For
simplicity and to make all calculations more transparent, we choose  $\Psi\Lb r'\Rb = K_0\Lb Q r'\Rb$.
The value of the parameter $Q$ can be  found from the value of  the electromagnetic radius of the proton ($Q\,\approx 0.3\,GeV^{-1}$).

In this presentation, we follow the formalism of Ref. \cite{LMKS}, namely
using the formulae {\bf 3.198},  {\bf 6.532(4)}, {\bf 6.565(4)}  and {\bf 6.566(2)} from the Gradstein and Ryzhik Tables, Ref. \cite{RY}. Introducing the Feynman parameter ($t$),  we can rewrite \eq{DP3} in the form
\bea \label{DP4}
&& \Psi\Lb r, z\Rb\,\,\,=\,\,\frac{\Gamma\Lb \Delta\Rb}{\pi\,\Gamma\Lb \Delta - 1\Rb}\,\int\,\xi\,d \xi\,d^2\, r' \frac{J_0\Lb Q\,\xi \Rb}{\xi^2\,+\,r'^2}\,\Lb \frac{z}{z^2\,\,+\,\,( \vec{r}\,-\,\vec{r}')^2}\Rb^{\Delta}\,\,=\,\frac{\Gamma\Lb \Delta + 1\Rb}{\pi\,\Gamma\Lb \Delta - 1\Rb}\,\frac{1}{B\Lb 1,\Delta\Rb}\nonumber\\
&& \times \,\,\int \xi\,d \xi\,d^2\, r'\int^1_0\,\frac{d t}{z}\,  t^{\Delta -1}\,(1 - t)\,\,J_0\Lb Q\,\xi \Rb\,\Lb \frac{z}{t\,z^2\,\,+\,\,t\,( \vec{r}\,-\,\vec{r}')^2\,\,+\,\,(1 - t)\,r'^2\,+\,(1 - t)\,\xi^2}\Rb^{\Delta + 1}\,\,\nonumber\\
 &&=\,\,\frac{\Gamma\Lb \Delta + 1\Rb}{\pi\,\Delta\,\Gamma\Lb \Delta - 1\Rb}\,\,\,z^{\Delta}\,\,\int \tilde{\xi}\,d \tilde{\xi}\,\int^1_0\,d t \, \frac{1}{(1 - t)^\Delta} \,\,J_0\Lb  Q\,\sqrt{\frac{t}{1 - t}}\,\tilde{\xi} \Rb\,\Lb \frac{1}{\,r^2\,\,\,+\,\,\kappa\Lb t,z,\tilde{\xi}\Rb}\Rb^{\Delta }
\eea
with $\kappa\Lb t,z,\xi\Rb \,=\,\Lb t\,z^2\,+\,\tilde{\xi}^2\Rb/\Lb  1 - t \Rb$ and $\tilde{\xi} = \xi\Lb\sqrt{1-t}/\sqrt{t}\Rb$.

The amplitude  $\tilde{A}_{1GE}\Lb s, b; z,z'\eq{N42}\Rb$ depends only on $z$ and $z'$, and  we need to find
$\int | \Psi\Lb r, z\Rb]^2 d^2 r$. From \eq{DP4}, one can see that we have to evaluate the integral
\bea \label{DP5}
&&\pi \int \,d r^2 \, \Lb\frac{1}{\,r^2\,\,\,+\,\,\kappa\Lb t,z,\tilde{\xi}\Rb}\Rb^{\Delta }\,\Lb \frac{1}{\,r^2\,\,\,+\,\,\kappa\Lb t',z,\tilde{\xi}'\Rb}\Rb^{\Delta }\,\,=\nonumber\\
&&\,\,\pi \,B\Lb 1, 2\Delta - 1\Rb {}_2F_1\Lb 1, \Delta, 2 \Delta - 1, 1 - \frac{\kappa\Lb t,z,\tilde{\xi}\Rb}{\kappa\Lb t',z,\tilde{\xi}'\Rb}\Rb \nonumber \\
&& \approx\,\,\pi\,\frac{1}{2 \Delta - 1}\,\frac{\kappa\Lb t,z,\tilde{\xi}\Rb}{\Lb \kappa\Lb t,z,\tilde{\xi}\Rb\,\kappa\Lb t',z,\tilde{\xi}'\Rb\Rb^{\Delta}}
\eea
where we used $ {\bf 3.197}$ of Ref. \cite{RY}.

In the last equation we assumed that $\kappa\Lb t,z,\xi\Rb/\kappa\Lb t',z,\xi'\Rb$ is close to unity, since
the integral has a symmetry with respect to $\xi \to xi'$, and $t \to t'$. The simplified form allows us
to reduce the integral for $\Phi(z)$ (see \eq{PHI}),  to the form

\bea \label{DP7}
\Phi\Lb z \Rb\,\,&=&\,\,z^{2 \Delta}\,\,\Lb \frac{\Gamma\Lb \Delta + 1\Rb}{\pi \Gamma\Lb \Delta - 1 \Rb}\Rb^2\,\,\frac{\pi}{2 \Delta - 1}\,\int\,\tilde{\xi}\, d \tilde{\xi}\,J_0\Lb Q\,\sqrt{\frac{t}{1 - t}}\,\tilde{\xi}\Rb\,
\,d\,t\,\,\frac{1}{\Lb t\,z^2\,+\,\,\tilde{\xi}^2\Rb^{\Delta }}\,\,\nonumber\\
&\times &
\int\,\tilde{\xi'}\, d \tilde{\xi}'\,J_0\Lb Q\,\sqrt{\frac{t'}{1 - t'}}\tilde{\xi}'\Rb\,\,d\,t'\,\,(1 - t')\,\frac{1}{\Lb t'\,z^2\,+\,(1 - t)\,\tilde{\xi}'^2\Rb^{\Delta - 1}}\,\,\nonumber\\
&=&\,\,\,\,\Lb \frac{2 \alpha'}{\rho}\Rb^{\Delta - 2}\,\,\frac{(\Delta -1)^2}{\Gamma\Lb \Delta\Rb\,\Gamma\Lb \Delta - 1 \Rb\,\pi}\,\,\frac{2\,Q^2\,2^{3 - 2 \Delta}\,z^3}{2 \Delta - 1}\,\,\,
\,\int^1_0 dt  \,\Lb \sqrt{\frac{t}{1 - t}} \,Q\Rb^{\Delta - 1}\,K_{\Delta - 1} \Lb \sqrt{\frac{t}{1 -t}}\,Q\,z\Rb\,\nonumber\\
&\times &\,\int^1_0 \,d t'\,\Lb \sqrt{\frac{t'}{1 - t'}} \,Q\Rb^{\Delta - 2}
K_{\Delta - 2} \Lb \sqrt{\frac{t'}{1 -t'}}\,Q\,z\Rb
\eea

 In the last equation we included the factor $\Lb \frac{2 \alpha'}{\rho}\Rb^{2 \Delta - 4}$, which recovers the correct dimension of the wave function. The origin of this factor is simple: we assumed for simplicity in all our previous calculations, that $L = 1$ in $AdS_5$. Since $L^2 = \alpha' \sqrt{\lambda} = \alpha' 2/\rho$, this factor is the way to take into account the fact that $L^2 \neq 1$.

\subsection{Qualitative features of high energy scattering}

From \eq{N46} one can see that $A_{NN}\Lb s,b\Rb$ tends to 1  in the region of $b$ from $b=0$ to $b =b_0(s)$. Since $G_2(u) \to 1/b^6$ at large $b$, we can conclude that $ A_{NN}\Lb s,b\Rb \,\,\propto s/b^6$ at $b \gg z^2$.
Therefore, $b^2_0 \,\propto s ^{1/3}$ and the nucleon-nucleon scattering amplitude generates the total cross section
\beq \label{QF1}
\sigma_{tot}\,\,=\,\,2 \int\,d^2 b\,Im  A_{NN}\Lb s,b\Rb\,\,\propto\,s^{1/3}
\eeq
in obvious violation of the Froissart theorem \cite{FROI}.  As has been shown in Refs. \cite{BST1,LMKS} the Froissart theorem can be restored if we consider the string theory which leads to N=4 SYM in the limit of the weak graviton interaction. In this string theory the graviton with positive $t$ ($t$ is the momentum transferred along the graviton in \fig{oge})  lies on the Regge trajectory with the intercept $\alpha'/2$ which corresponds to the closed string. On the other hand in $AdS_5$ the Einstein equation has the form
\beq \label{EE}
R_{\mu,\nu}\,\,-\,\,\frac{1}{2} \,R\,g_{\mu,\nu}\,=\,\,\frac{6}{L^2}\,\,g_{\mu,\nu}\,\,\,=\,\,\frac{6}{\sqrt{\lambda}\,\alpha'}\,g_{\mu,\nu}
\eeq
where $R_{\mu,\nu}$ is the Ricci curvature tensor and $R$ is the Ricci curvature.  In consequence of \eq{EE} the graviton has the mass\cite{GMA}
$m^2_{graviton} \,=\,4/\sqrt{\lambda} \alpha'\,=\,2\,\rho/\alpha'$ and the intercept $2 - m^2_{graviton} (\alpha'/2)\,=\,2 - 2/\sqrt{\lambda}\,\,=\,\,2 - \rho$.

The fact that the graviton has mass results in the different behavior of the gluon propagator at large $b$, namely,    at large $b$  it shows the exponential decrease
$G(b) \to \exp\Lb - m_{graviton}\,b\Rb  = \exp\Lb - \sqrt{2\,\rho\,b^2} \Rb$.
 Such behavior restores  the logarithmic dependence of the cross section at high energy, in the agreement with the Froissart theorem but nevertheless we expect a wide range of energies where the cross section behaves as $s^{1/3}$. Experimentally, the total cross section in the energy range from fixed target experiment at FNAL to the Tevatron energy, has $\sigma_{tot} \propto s^{0.1}$.
Therefore, we expect that the cross section cannot be described by \eq{N46}.

We replace $G_3(u) $ in \eq{N46} and in \eq{EQ} by
\beq \label{QF2}
 \tilde{A}_{1GE}(s,b; z,z')\,\,\,\longrightarrow\,\,\,\,\tilde{A}_{1GE}(s,b; z,z')\,\,e^{- \sqrt{2\rho/\alpha'}\,b}
\eeq
to take into account the effect of the graviton reggeization. Introducing this equation we are able to specify the kinematic energy range, where we expect the $s^{1/3}$ behavior of the total cross section.

\section{The comparison with the experimental data}
As we have discussed, we face two main difficulties in our attempts to describe the experimental data in N=4 SYM: the small value of the cross section of the multiparticle  production and the violation of the Froissart theorem.  The scale of both phenomena is given by  the value of $2/\sqrt{\lambda}$ (see \eq{N45} and \eq{QF2})  and, if this parameter is not small, we, perhaps, have no difficulties at all.  On the other hand, the N=4 SYM could provide the educated guide only for  $2/\sqrt{\lambda}\,\ll\,1$ since it has an analytical solution for such $\lambda$. Therefore, the goal of our approach is  to describe the experimental data assuming that $2/\sqrt{\lambda}\,$ is reasonably small (say $2/\sqrt{\lambda}\,\leq\,0.3$), and to evaluate the scale of the cross section for the multiparticle production. As has been mentioned, the multiparticle production
 can be discussed in N=4 SYM since the confinement of the quarks and gluon, we believe , is not essential for these processes.

We use \eq{N45} with \eq{DP7} to calculate the physical observables, namely,
\bea
\sigma_{tot}\,\,&=&\,\sigma_0\,+\,\,2 \,\int d^2 b \, \,Im\,A\Lb s,b\Rb\,\,\label{EX11}\\
&=&\,\,\sigma_0\,+\,\frac{4}{\rho \alpha'}\,\int d^2 b \,\,\int \Phi(z)\,\Phi(z')\,d z \,d z'
\,Re \left\{ 1  \,\,-\,\,\exp\Lb i N^2_c\,\tilde{A}_{1GE}\Lb s,b,z,z'|\eq{EQ}\Rb\Rb\,\right\}\,\nonumber\\
\sigma_{el} &=& \int d^2 b  \,\,| A_0(b)\,+\,A\Lb s,b\Rb|^2\,\,\label{EX12}\\
 &=&\,\,\int d^2 b \,\, | A_0(b)\,+\,\int \Phi(z)\,\Phi(z') \,d z\, d z'
\,i\left\{ 1  \,\,-\,\,\exp\Lb i N^2_c\,\tilde{A}_{1GE}\Lb s,b,z,z'|\eq{EQ}\Rb\Rb\,\right\}|^2;\nonumber\\
B_{el} &=&\, \,\frac{\int d^2 b\, \,b^2 \,\,| A_0(b)\,+\,\int \Phi(z)\,\Phi(z') \,d z\, d z'
\,i\left\{ 1  \,\,-\,\,\exp\Lb i N^2_c\,\tilde{A}_{1GE}\Lb s,b,z,z'|\eq{EQ}\Rb\Rb\,\right\}|^2}{\int d^2 b \,\, | A_0(b)\,+\,\int \Phi(z)\,\Phi(z') \,d z\, d z'
\,i\left\{ 1  \,\,-\,\,\exp\Lb i N^2_c\,\tilde{A}_{1GE}\Lb s,b,z,z'|\eq{EQ}\Rb\Rb\,\right\}|^2} \,;\label{EX13}\,\,
\eea

As has been expected, it turns out that in the experimental accessible region  of energies, the cross section given in \eq{N46} shows the
$s^{1/3}$ behavior for a wide range of parameters: $ g^2 = 0.01 \div 1$, $Q = 0.2 \div 1 \,GeV^{-1}$ and $\rho = 0 \div 0.3$.  Our choice of the parameters reflects the theoretical requirements for N=4 SYM, where we can trust
this approach, namely, $g_s  \ll 1$ while $g_s\,N_c > 1$.
The values of $\sigma_{tot}$ from \eq{N46} with $\tilde{A}$ from \eq{EQ} are small for $W = \sqrt{s}= 20 GeV$ but it increases and becomes about $20-30\,mb$ at the Tevatron energy.  Facing the clear indication that we need an extra contribution to the total cross section in \eq{EX11}-\eq{EX13},  we introduce the  contribution of the non N=4 SYM origin ($\sigma_0 $ and  the amplitude $A_0(b)$ ).

It should be mentioned that we have also a hidden parameter $\Delta$ in the wave function of the proton.  At the moment theoretically we know only that $\Delta > 2$. This constraint stems from the convergence of the integral for the norm of the proton wave function (see Refs.\cite{WIT,MHI,BEPI,GRF4M}).  We have tried several values of $\Delta$ and $\Delta =3$ is our best choice (see \fig{dep2}).

For a purely phenomenological background $A_0(b)$ we wrote the simplest expression
\beq \label{EX2}
A_0(b)\,\,=\,\,i\,\frac{\sigma_0}{4\,\pi\, B_0}\,\exp\Lb - b^2/2 B_0\Rb
\eeq
 where $B_0$ is the slope for the elastic cross section.

With these two new parameters $\sigma_0$ and $ B_0$, we tried to describe the data. The results are shown in \fig{sig},\fig{bel} and \fig{sel}.

\FIGURE[t]{
\centerline{\epsfig{file=w_1_tot.epsi,width=100mm,height=70mm}}
\vspace{-0.5cm}
\caption{The description of the total cross section $\sigma_{tot} =(\sigma_{tot}( p p )  + \sigma_{tot}(p \bar{p}))/2$
   \,with $Q= 0.35 GeV$, $g\,= \,g^2_s\,N^2_c = 0.1$, $\rho = 0.25$ , $\Delta =3$ and with $\sigma_0 = 37.3\,mb$.
\label{sig} }}

From these pictures one can see that for
the total and elastic cross section,   we obtain a good agreement with the experimental data, whereas for the elastic slope ($B_{el}$), the description is in contradiction with the experimental data.  First we would like to understand the main ingredients of the total cross section. For doing so we need to estimate the cross section of the diffractive dissociation. In the N=4 SYM  approach;

\bea \label{DD}
\sigma_{diff}\,\,&=&\,\, \frac{2}{\rho \alpha'}\,\int d^2 b \,\,\int \Phi(z)\,\Phi(z')\,d z \,d z'
\,|  1  \,\,-\,\,\exp\Lb i N^2_c\,\tilde{A}_{1GE}\Lb s,b,z,z'|\eq{EQ}\Rb\Rb|^2\nonumber\\
 & - &\,\,|\frac{2}{\rho \alpha'}\,\int d^2 b \,\,\int \Phi(z)\,\Phi(z')\,d z \,d z'
\,|  1  \,\,-\,\,\exp\Lb i N^2_c\,\tilde{A}_{1GE}\Lb s,b,z,z'|\eq{EQ}\Rb\Rb||^2
\eea

In \eq{DD} $\sigma_{diff} = \sigma_{sd} + \sigma_{dd}$ where $\sigma_{sd}$ and $\sigma_{dd}$ are cross sections of single and double diffraction respectively. Our predictions for $\sigma_{diff}$ have been plotted in \fig{sdd}, where curve 1 is the result of the calculation using \eq{DD}, and curve 2 is the same except for the addition of $4 mb$ from the diffractive cross section, which is of non N=4 SYM origin. In Table\ref{t}, we compare our predictions with the phenomenological models that do not take into account the N=4 SYM physics. The result of this comparison is interesting, since our simple estimates show that the cross section of the diffractive production, could considerably grow from the Tevatron to the LHC energy. We want to recall that the unitarity constraints of \eq{N45}, lead to $|A(s,b;z,z')| \,\leq 2$ and $\sigma_{tot} = \sigma_{el}$.

As far as the inelastic cross section is concerned, one can see that the inelastic cross section of the N=4 SYM origin
$\sigma\Lb\mbox{N=4 SYM}\Rb\,\,=\,\,\sigma_{tot} - \sigma_{el} - \sigma_{diff} - \sigma_{0,in}$ is about 2 $mb $
both for RHIC and the Tevatron energy, and grows to  30  $mb$ at the LHC energy. Therefore, we can  observe some typical features of the N=4 SYM theory, which only start at the LHC energy.

The above estimates    are based on the background that does not depend on energy.  However, \fig{bel} illustrates that the non N=4 SYM background should also depend on energy. In \fig{bel} (the upper curve) we plot the elastic slope for the background of \eq{EX2} but with $B_0 = 12.37 + 2 \alpha'_P \ln(s/s_0)$. This amplitude corresponds to the exchange of the Pomeron with intercept 1 which generates the constant cross section but leads to a shrinkage of the diffraction peak. One can see that we are able to describe the slope in such a model.

\FIGURE[t]{
\centerline{\epsfig{file=w_1_slope.epsi,width=100mm,height=70mm}}
\caption{The description of the energy behavior of the elastic slope  with the same set of parameters as in \protect\fig{sig} and with $B_0 = 12.37 \,GeV^{-2}$ (solid curve) and  $B_0 = 12.37 + 2\, \alpha'_P\ln(s/s_0)$ ($\alpha'_P=0.1\,GeV^{-2}$(dashed  curve) and $\alpha'_P=0.2\,GeV^{-2}$ (dotted curve))
\label{bel} }}

\FIGURE[t]{
\centerline{\epsfig{file=w_1_el.epsi,width=100mm,height=70mm}}
\caption{The description of the energy behavior of the elastic cross section  with the same set of parameters as in \protect\fig{sig} and $B_0 = 12.37\,GeV^{-2}$.}
\label{sel} }

\TABLE[ht]{
\begin{tabular}{||l|l|l||}
\hline \hline
  &  \,\,\,\,\,\,\,\,\,\,\,\,\,\,\,\,\,\,\,\,\,\,\,\,Tevatron &
\,\,\,\,\,\,\,\,\,\,\,\,\,\,\,\,\,\,\,\,\,\,\,\, LHC \\
 & GLMM\,\,\,\,\,\,\,\,\,\,\,\,\,\,\,\,\,\,\,\,\,\,KMR\,\,\,\,\,\,\,\,\,\,\,\,\,\,\,\,\,\,\,\,\,LP &
GLMM\,\,\,\,\,\,\,\,\,\,\,\,\,\,\,\,\,\,\,\,KMR \,\,\,\,\,\,\,\,\,\,\,\,\,\,\,\,\,\,\,LP  \\\hline
$\sigma_{tot}$( mb ) & 73.29 \,\,\,\,\,\,\,\,\,\,\,\,\,\,\,\,\,\,\,\,\,\,\,\,74.0 \,\,\,\,\,\,\,\,\,\,\,\,\,\,\,\,\,\,\,\,\,\,\,\,83.2
& 92.1\,\,\,\,\,\,\,\,\,\,\,\,\,\,\,\,\,\,\,\,\,\,\,\,\,\,88.0 \,\,\,\,\,\,\,\,\,\,\,\,\,\,\,\,\,\,\,\,\,\,\,\,\,124.9
 \\ \hline
$\sigma_{el}$(mb) & 16.3 \,\,\,\,\,\,\,\,\,\,\,\,\,\,\,\,\,\,\,\,\,\,\,\,\,\,\,16.3 \,\,\,\,\,\,\,\,\,\,\,\,\,\,\,\,\,\,\,\,\,\,\,\,\,17.5 &
20.9 \,\,\,\,\,\,\,\,\,\,\,\,\,\,\,\,\,\,\,\,\,\,\,\,20.1\,\,\,\,\,\,\,\,\,\,\,\,\,\,\,\,\,\,\,\,\,\,\,\,\,\,\,24.4 \\\hline
$\sigma_{sd} \,+\,\sigma_{dd}$(mb) & 15.2 \,\,\,\,\,\,\,\,\,\,\,\,\,\,\,\,\,\,\,\,\,\,\,\,\,\,18.1\,  \,\,\,\,\,\,\,\,\,\,\,\,\,\,\,\,\,\,\,\,\,\,\,\,\,24.4 &
17.88 \,\,\,\,\,\,\,\,\,\,\,\,\,\,\,\,\,\,\,\,\,\,26.7 \,\,\,\,\,\,\,\,\,\,\,\,\,\,\,\,\,\,\,\,\,\,\,42.3\\\hline
$\Lb\sigma_{el} + \sigma_{sd} + \sigma_{dd}\Rb/\sigma_{tot}$ &
0.428 \,\,\,\,\,\,\,\,\,\,\,\,\,\,\,\,\,\,\,\,\,\,\,\,0.464\,\,\,\,\,\,\,\,\,\,\,\,\,\,\,\,\,\,\,\,\,\,\,\,0.504 &
0.421 \,\,\,\,\,\,\,\,\,\,\,\,\,\,\,\,\,\,\,\,\,\,0.531\,\,\,\,\,\,\,\,\,\,\,\,\,\,\,\,\,\,\,\,\,\,0.536 \\
\hline \hline
\end{tabular}
\caption{Comparison of the GLMM (\cite{GLMM}) and  KMR\cite{KMRNEW} models and our estimates (LP).
}
\label{t}}

In \fig{dep1} and \fig{dep2} we plot the dependence of $\sigma_{tot}$ and $\sigma_{el}$ on the parameters of our approach to illustrate the sensitivity of our descriptions of the experimental data to their values.
\FIGURE[t]{
\centerline{\epsfig{file=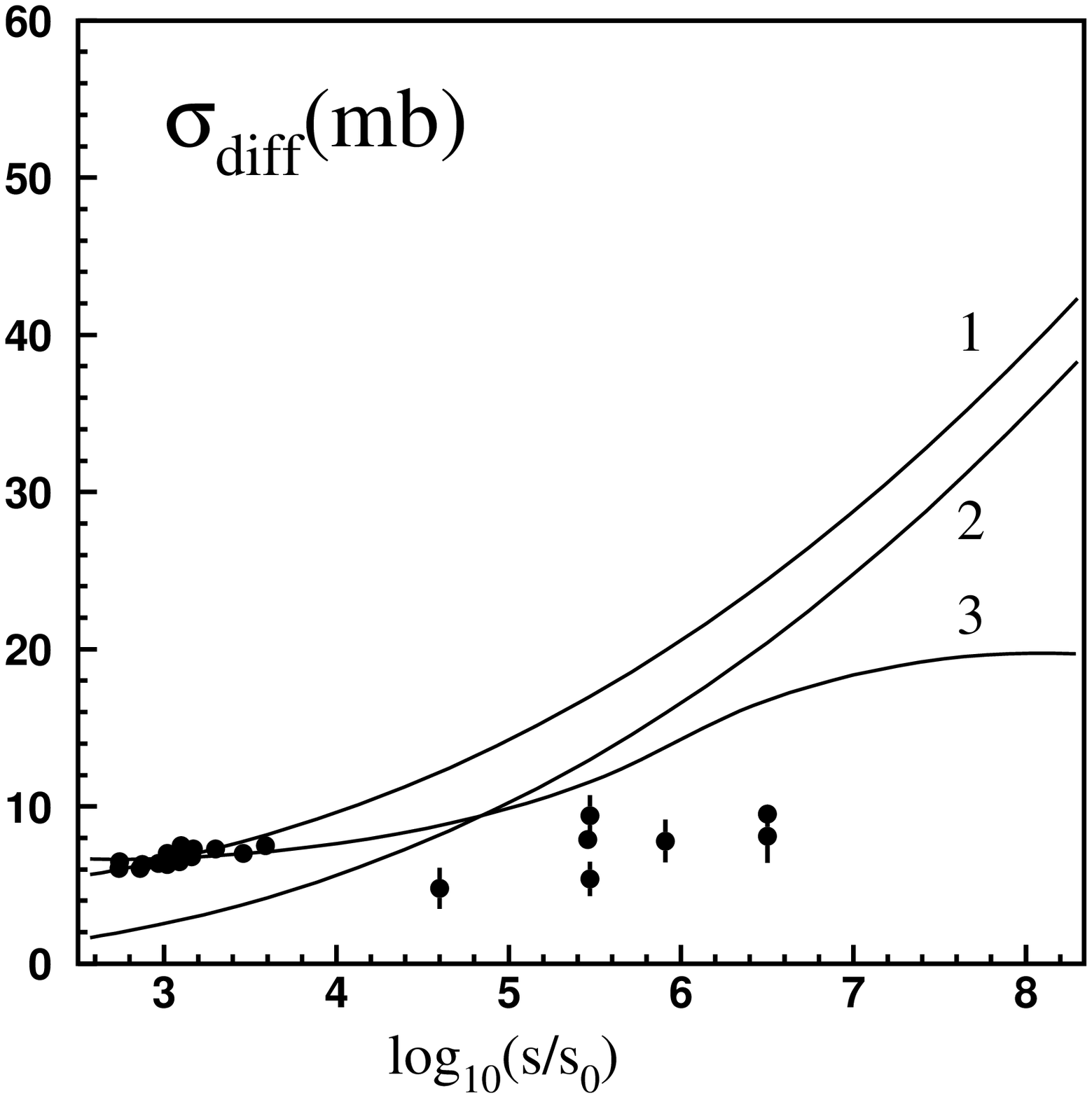,width=100mm,height=70mm}}
\caption{The description of the energy behavior of the diffraction production cross section $\sigma_{diff} \,=\,\sigma_{sd} \,+\,\sigma_{dd}$  with the same set of parameters as in \protect\fig{sel}.
$\sigma_{sd} $ and $\sigma_{dd} $ are cross sections of  single and double diffraction production respectively.
 The curve 2 shows the N=4 SYM contribution to the diffraction production while  the curve 1 corresponds to the N=4 SYM prediction plus $4\,mb$ for the cross section of a different source than N=4 SYM. The data are only for single diffraction production.  In curve 3 we plot the estimates of Ref.\cite{GLMM} for $\sigma_{diff}$
\label{sdd} }}

\FIGURE[t]{
\centerline{\epsfig{file=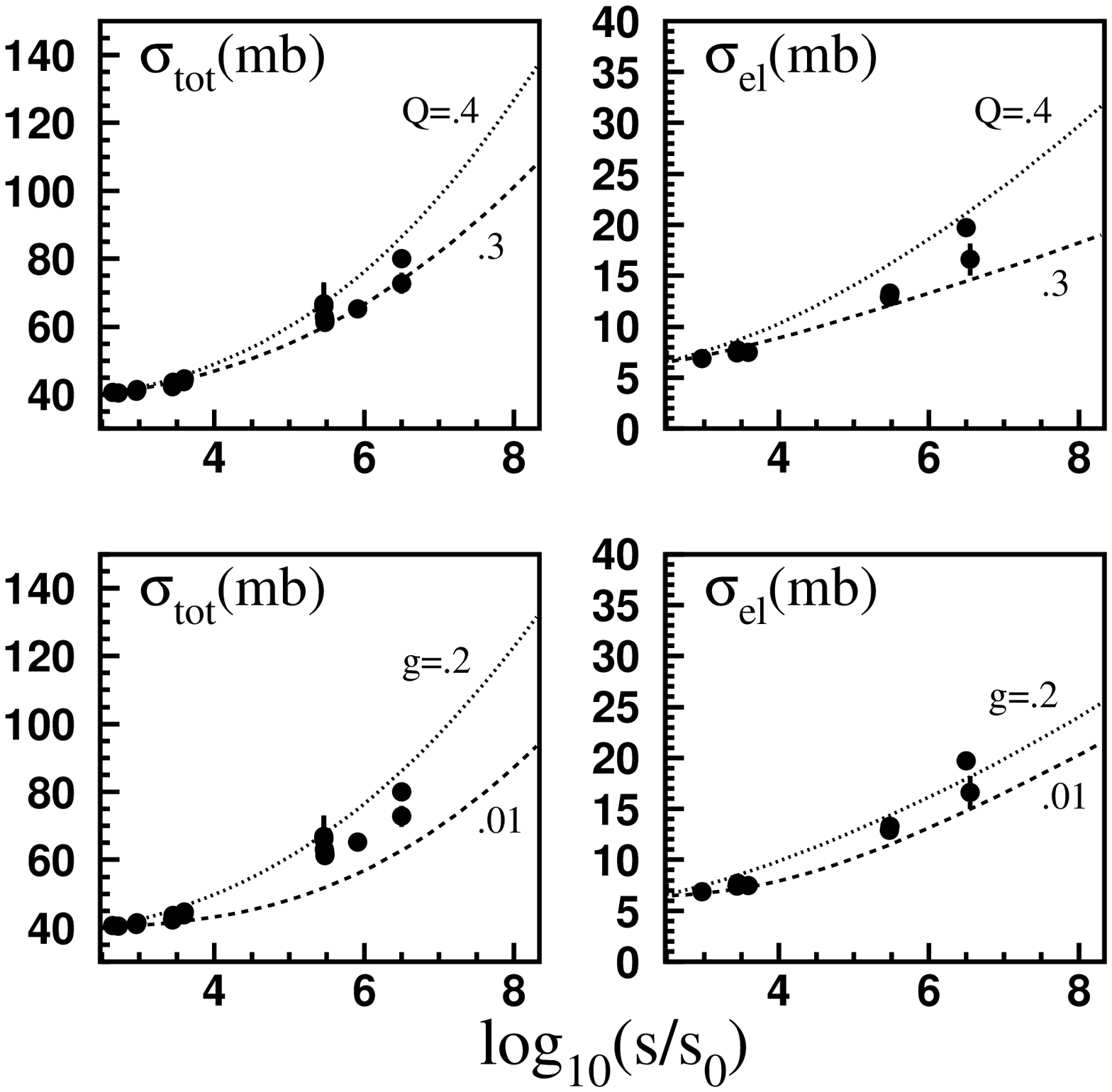,width=110mm,height=90mm}}
\caption{The dependence of the description on $Q$ and $g=N^2_c\,g^2$.
\label{dep1} }}
 \FIGURE[t]{
\centerline{\epsfig{file=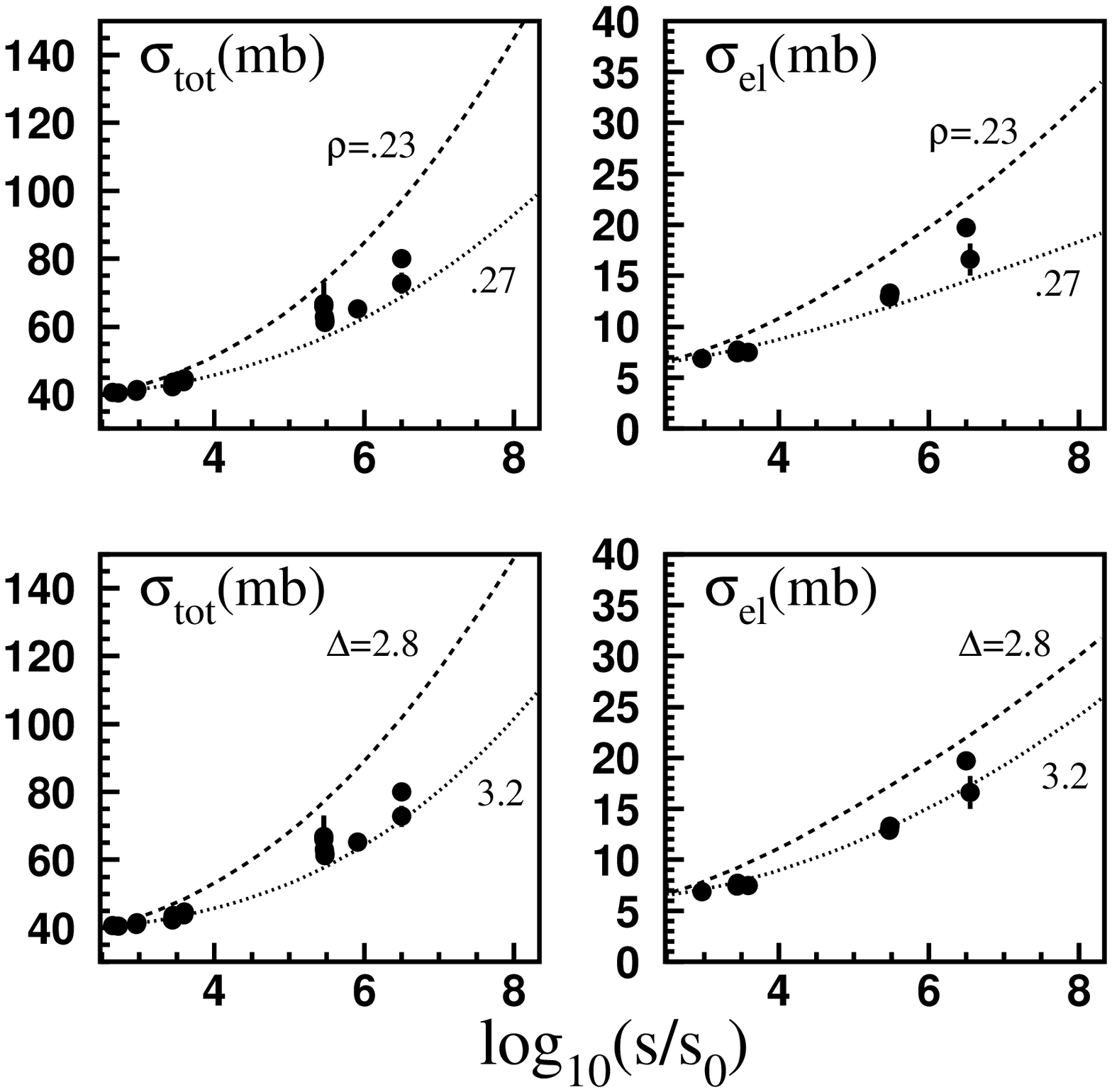,width=100mm,height=80mm}}
\caption{The dependence of the description on $\rho$ and $\Delta$.
\label{dep2} }}

The results of our calculation show that
in the large range of energies, the N=4 SYM scattering amplitude behaves as $s^{1/3}$ with a rather small coefficient in front. The graviton reggeization that will stop the anti-Froissart behavior at ultra high energies, does not show up at the accessible range of energy from the fixed target Fermilab energy, until the Tevatron  energy.  This reggeization can be measured, perhaps,  only     at the LHC energy.

\section{Conclusion}
In this paper we show that the N=4 SYM total cross section violates the Froissart theorem, and in the huge range of energy this cross section is proportional to $s^{1/3}$. The graviton reggeization will change this increase to the normal logarithmic behavior $\sigma \propto \ln^2 s$.  However,  we demonstrated that this happens at  ultra high energy, much higher than the LHC energy  for reasonably  low $2/\sqrt{\lambda} \approx 0.25$.

We need to assume that there is a  different source for the total cross section, with the value of the cross section about 40 mb.  With this assumption we successfully describe $\sigma_{tot}, \sigma_{el}$ and $\sigma_{diff}$
for the accessible range of energy from the fixed target Fermilab to the Tevatron energies.  The N=4 SYM mechanism is responsible only for a small part of the inelastic cross section for this energy region (about $2 mb$).
However, at the LHC energy the N=4 SYM theory can lead to  a valuable contribution to the inelastic cross section, namely,  $\sigma_{in} \approx  30 \,mb$ which is about a quarter of the total inelastic cross section.
The second surprise is the fact that the total cross section and the diffraction cross section can increase considerably from the Tevatron to the LHC energy. The bad description of $B_{el}$ gives the strong argument that the non N=4 SYM background should depend on energy.

 It means that  at RHIC energies, the N=4 SYM part of the inelastic cross section is negligible and the quark-gluon plasma is created by the mechanism outside of N=4 SYM. For the LHC energy, we can expect that N=4 SYM  is responsible for the inelastic cross section of about $\sigma_{in}(N=4\,\,SYM) =30\,\, mb$ out of $\sigma_{tot} =121.9\,mb$.

We believe that we have a dilemma: to find a new mechanism for the inelastic production in the framework of N=4 SYM other than  the reggeized graviton interaction, or to accept that N=4 SYM is irrelevant to description of  any  experimental data that have been  measured  before the LHC era, with a chance that even at the LHC it will be responsible only for a quarter   (or less) of the total cross section. Deeply in our hearts, we believe in the first way out, and we hope that this paper will draw attention to this challenging problem: searching for a new mechanism for multiparticle production in N=4 SYM.

We wish to draw your attention to the fact that the scattering amplitude can change considerably from the Tevatron to LHC energy (see Table 1). Therefore, all claims that we can  give reliable predictions for   the  values of the cross sections at the LHC energy and even of  the survival probability for the diffractive Higgs production \cite{KMRNEW}  looks exclusively naive and reflects our prejudice rather than our understanding.

\section* {Acknowledgments}
We thank Boris Koppeliovich for fruitful discussion on the subject of the  paper. Our special thanks go to Miguel Costa and Jeremy Miller  for their  careful reading of the first version of this paper and useful discussions.
E.L.
also thanks the  high energy theory group of the University Federico
Santa Maria for the hospitality and creative atmosphere during his visit.

 This work was supported in part by Fondecyt (Chile) grants, numbers 1050589,
7080067 and 7080071, by DFG (Germany)  grant PI182/3-1 and
 by BSF grant $\#$ 20004019.


\begin{thebibliography}{99}
\bibitem{AdS-CFT}
 J.~M.~Maldacena,
  Adv.\ Theor.\ Math.\ Phys.\  {\bf 2} (1998) 231
  [Int.\ J.\ Theor.\ Phys.\  {\bf 38} (1999) 1113]
  [arXiv:hep-th/9711200];\,\,\,
S.~S.~Gubser, I.~R.~Klebanov and A.~M.~Polyakov,
  Phys.\ Lett.\  B {\bf 428} (1998) 105
  [arXiv:hep-th/9802109];\,\,\,
E.~Witten,
  Adv.\ Theor.\ Math.\ Phys.\  {\bf 2} (1998) 505
  [arXiv:hep-th/9803131].
\bibitem{POST}
 J.~Polchinski and M.~J.~Strassler,
  JHEP {\bf 0305} (2003) 012
  [arXiv:hep-th/0209211];\,\,
  Phys.\ Rev.\ Lett.\  {\bf 88} (2002) 031601
  [arXiv:hep-th/0109174].
\bibitem{BFKL4}
 A.~V.~Kotikov, L.~N.~Lipatov, A.~I.~Onishchenko and V.~N.~Velizhanin,
  Phys.\ Lett.\  B {\bf 595} (2004) 521
  [Erratum-ibid.\  B {\bf 632} (2006) 754]
  [arXiv:hep-th/0404092];\,\,
A.~V.~Kotikov and L.~N.~Lipatov,
  Nucl.\ Phys.\  B {\bf 661} (2003) 19
  [Erratum-ibid.\  B {\bf 685} (2004) 405]
  [arXiv:hep-ph/0208220];\,\,
  A.~V.~Kotikov and L.~N.~Lipatov,
  Nucl.\ Phys.\  B {\bf 582} (2000) 19
  [arXiv:hep-ph/0004008].
  A.~V.~Kotikov, L.~N.~Lipatov and V.~N.~Velizhanin,
  Phys.\ Lett.\  B {\bf 557} (2003) 114
  [arXiv:hep-ph/0301021];\,\,\,J.~R.~Andersen and A.~Sabio Vera,
  Nucl.\ Phys.\  B {\bf 699} (2004) 90
  [arXiv:hep-th/0406009];\,\,\,\,Z.~Bern, M.~Czakon, L.~J.~Dixon, D.~A.~Kosower and V.~A.~Smirnov,
  Phys.\ Rev.\  D {\bf 75} (2007) 085010
  [arXiv:hep-th/0610248];\,\,  Z.~Bern, L.~J.~Dixon and V.~A.~Smirnov, Phys.\ Rev.\  D {\bf 72} (2005) 085001
  [arXiv:hep-th/0505205];\,\,
\bibitem{MHI}
Y.~Hatta, E.~Iancu and A.~H.~Mueller,
  JHEP {\bf 0801} (2008) 026
  [arXiv:0710.2148 [hep-th]].




\bibitem{BST1}
 R.~C.~Brower, J.~Polchinski, M.~J.~Strassler and C.~I.~Tan,
  JHEP {\bf 0712} (2007) 005
  [arXiv:hep-th/0603115].
\bibitem{BST2}
 R.~C.~Brower, M.~J.~Strassler and C.~I.~Tan,
  arXiv:0707.2408 [hep-th].
\bibitem{BST3}
 R.~C.~Brower, M.~J.~Strassler and C.~I.~Tan,
  JHEP {\bf 0806} (2008) 048
  [arXiv:0801.3002 [hep-th]].
  \bibitem{COCO}
 L.~Cornalba and M.~S.~Costa,
 Phys. Rev. {\bf D 78}, (2008) 09010,
  arXiv:0804.1562 [hep-ph];\,\,\,
  L.~Cornalba, M.~S.~Costa and J.~Penedones,
  JHEP {\bf 0806} (2008) 048
  [arXiv:0801.3002 [hep-th]];\,\,
  JHEP {\bf 0709} (2007) 037
  [arXiv:0707.0120 [hep-th]].

\bibitem{BEPI}
B.~Pire, C.~Roiesnel, L.~Szymanowski and S.~Wallon,
  Phys.\ Lett.\  B {\bf 670}, 84 (2008)
  [arXiv:0805.4346 [hep-ph]].
\bibitem{LMKS}
E.~Levin, J.~Miller, B.~Z.~Kopeliovich and I.~Schmidt,
JHEP {\bf 0902} (2009) 048;\,\,
  arXiv:0811.3586 [hep-ph].
\bibitem{GLR}
L. V. Gribov, E. M. Levin and M. G. Ryskin, {\it Phys. Rep.}\,
{\bf 100}, 1 (1983).

\bibitem{MUQI}
A. H. Mueller and J. Qiu,  {\it Nucl. Phys.} \,{\bf B 268}\,427
(1986) .
\bibitem{MV}
L. McLerran and R. Venugopalan, {\it  Phys. Rev.}  {\bf D 49},2233,
3352  (1994); {\bf D 50},2225 (1994); {\bf D 53},458 (1996); {\bf
D 59},09400
(1999).


\bibitem{BFKL}
 E. A. Kuraev, L. N. Lipatov, and F. S. Fadin, {\it  Sov. Phys.
JETP}
                {\bf 45}, 199 (1977); \,\,\,
Ya. Ya. Balitsky and L. N. Lipatov,
               {\it   Sov. J. Nucl. Phys.}\, {\bf 28}, 22 (1978).

\bibitem{BK}
I.~Balitsky,
[arXiv:hep-ph/9509348];\,\,
{\it Phys.\ Rev.} {\bf D60}, 014020 (1999)
[arXiv:hep-ph/9812311]\,\,\,\,
Y.~V.~Kovchegov,
{\it Phys.\ Rev.}  {\bf D60}, 034008  (1999),
[arXiv:hep-ph/9901281].
\bibitem{JIMWLK}
~J.~Jalilian-Marian, A.~Kovner, A.~Leonidov and H.~Weigert,
{\it  Phys.\ Rev.}\,  {\bf D59}, 014014 (1999),
[arXiv:hep-ph/9706377];\,\,  {\it Nucl.\ Phys.}\,{\bf B504}, 415
(1997),
[arXiv:hep-ph/9701284]; \,\,\,
J.~Jalilian-Marian, A.~Kovner and H.~Weigert,
  {\it Phys.\ Rev.}  {\bf D59}, 014015 (1999),
  [arXiv:hep-ph/9709432];\,\,\,
 A.~Kovner, J.~G.~Milhano and H.~Weigert,
 {\it  Phys.\ Rev.}  {\bf D62}, 114005 (2000),
  [arXiv:hep-ph/0004014]\,; \,\,\,
E.~Iancu, A.~Leonidov and L.~D.~McLerran,
{\it  Phys.\ Lett.}\,  {\bf B510}, 133 (2001);
[arXiv:hep-ph/0102009];\,\, {\it  Nucl.\ Phys.}\,  {\bf A692}, 583
(2001),
[arXiv:hep-ph/0011241];\,\,\,
E.~Ferreiro, E.~Iancu, A.~Leonidov and L.~McLerran,
 {\it  Nucl.\ Phys.}\  {\bf A703}, 489 (2002),
  [arXiv:hep-ph/0109115];\,\,\,
H.~Weigert,
{\it  Nucl.\ Phys.}  {\bf A703}, 823 (2002),
[arXiv:hep-ph/0004044].


\bibitem{KSS}
 P.~Kovtun, D.~T.~Son and A.~O.~Starinets,
  Phys.\ Rev.\ Lett.\  {\bf 94} (2005) 111601
  [arXiv:hep-th/0405231].
\bibitem{HKKKY}
C.~P.~Herzog, A.~Karch, P.~Kovtun, C.~Kozcaz and L.~G.~Yaffe,
  JHEP {\bf 0607} (2006) 013
  [arXiv:hep-th/0605158].
\bibitem{MUN4}
 A.~H.~Mueller,
  Phys.\ Lett.\  B {\bf 668} (2008) 11
  [arXiv:0805.3140 [hep-ph]];\,\,\,
  F.~Dominguez, C.~Marquet, A.~H.~Mueller, B.~Wu and B.~W.~Xiao,
  Nucl.\ Phys.\  A {\bf 811} (2008) 197
  [arXiv:0803.3234 [nucl-th]];\,\,\,
 Y.~Hatta, E.~Iancu and A.~H.~Mueller,
  JHEP {\bf 0805} (2008) 037
  [arXiv:0803.2481 [hep-th]]\;\,\,\,
\bibitem{PLASMAN4}
P.~M.~Chesler and L.~G.~Yaffe,
  Phys.\ Rev.\  D {\bf 78} (2008) 045013
  [arXiv:0712.0050 [hep-th]];\,\,\,
A.~Yarom,
  Phys.\ Rev.\  D {\bf 75} (2007) 105023
  [arXiv:hep-th/0703095].



\bibitem{WIT}
E.~Witten,
  Adv.\ Theor.\ Math.\ Phys.\  {\bf 2} (1998) 253
  [arXiv:hep-th/9802150].


\bibitem{RY}
I. Gradstein and I. Ryzhik, {\it `` Tables of Series, Products, and
Integrals"}, Verlag MIR, Moskau,1981


\bibitem{FROI}
M.~Froissart,
{\it Phys.\, Rev.} \,  {\bf 123} (1961) 1053; \\
~A. ~Martin, {``Scattering Theory: Unitarity, Analitysity and Crossing."}
Lecture Notes in Physics, Springer-Verlag,  Berlin-Heidelberg-New-York,
1969.

\bibitem{GMA}
 J.~Naf, P.~Jetzer and M.~Sereno,
  Phys.\ Rev.\  D {\bf 79} (2009) 024014
  [arXiv:0810.5426 [astro-ph]]\,;\,\,\,\,L.~Liu,
  arXiv:gr-qc/0411122\,;M.~Novello and R.~P.~Neves,
{\it Class. Quantum  Grav.} {\bf 20} (2003) 67,
  arXiv:gr-qc/0210058.
\bibitem{GRF4M}
E.~D'Hoker, D.~Z.~Freedman, S.~D.~Mathur, A.~Matusis and L.~Rastelli,
  Nucl.\ Phys.\  B {\bf 562} (1999) 330
  [arXiv:hep-th/9902042];\,\,\,
E.~D'Hoker and D.~Z.~Freedman,
  Nucl.\ Phys.\  B {\bf 550} (1999) 261
  [arXiv:hep-th/9811257];\,\,\,
D.~Z.~Freedman, S.~D.~Mathur, A.~Matusis and L.~Rastelli,
  Nucl.\ Phys.\  B {\bf 546} (1999) 96
  [arXiv:hep-th/9804058];\,\,\,
 E.~D'Hoker, D.~Z.~Freedman, S.~D.~Mathur, A.~Matusis and L.~Rastelli,
  Nucl.\ Phys.\  B {\bf 562} (1999) 353
  [arXiv:hep-th/9903196];\,\,\,
H.~Liu,
  Phys.\ Rev.\  D {\bf 60} (1999) 106005
  [arXiv:hep-th/9811152].

\bibitem{GLMM}
 E.~Gotsman, E.~Levin, U.~Maor and J.~S.~Miller,
  Eur.\ Phys.\ J.\  C {\bf 57} (2008) 689
  [arXiv:0805.2799 [hep-ph]].
\bibitem{KMRNEW}
M. G. Ryskin, A. D. Martin and V. A. Khoze,
Eur. Phys. J. {\bf C54} (2008) 199 [arXiv:0710.2494 [hep-ph]].
\end{thebibliography}
\end{document}